\documentclass[useAMS,usenatbib]{mn2e}

\usepackage{natbib}
\usepackage{color}
\usepackage{amsmath, amssymb}
\usepackage{bm}
\usepackage{pifont}
\usepackage{float}
\usepackage{graphicx}
\usepackage[usenames,dvipsnames]{xcolor}
 \usepackage{relsize}
\usepackage{subfigure}  %inclusion de varias imagenes
\usepackage[singlelinecheck=false % <-- important
]{caption}

\usepackage{soul} %strikeoutcommand \st{text}
\usepackage{epsfig}
\usepackage{enumerate}
\usepackage[shortlabels]{enumitem}
\usepackage[version=3]{mhchem}
\usepackage{caption}
\usepackage{varwidth}
\newsavebox\CBox
\usepackage{color}
\usepackage{lineno}
\usepackage{epstopdf} % with I can solve the problem about the pictures.
\usepackage{float}

\newcommand{\sed}[1]{{\textcolor{black}{#1}}}

\DeclareMathOperator\arctanh{arctanh}

\title[Ionisation and discharge  in cloud-forming atmospheres of brown dwarfs and extrasolar planets]{Ionisation and discharge in cloud-forming atmospheres of brown dwarfs and extrasolar planets}

\author[Ch. Helling, P. B. Rimmer, I. M. Rodriguez-Barrera, K. Wood, G.B. Robertson, C.R. Stark]{{Ch. Helling$^1$\thanks{E-mail: ch80@st-andrews.ac.uk}, P. B. Rimmer$^1$, I. M. Rodriguez-Barrera$^1$, Kenneth Wood$^1$,}
\newauthor{G.B. Robertson$^1$, C.R. Stark$^2$ }\\
$^1$ SUPA, School of Physics and Astronomy, University of St Andrews, St Andrews KY16 9SS, UK\\
$^2$ Division of Computing and Mathematics, School of Arts, Media and Computer Games, Abertay University, Dundee DD1 1HG, UK.
}

\begin{document} 
\maketitle

%\linenumbers
\begin{abstract}
Brown dwarfs and giant gas extrasolar planets have cold atmospheres
with a rich chemical compositions from which mineral cloud particles
form. Their properties, like particle sizes and material
  composition, vary with height, and the mineral cloud particles are
  charged due to triboelectric processes in such dynamic
  atmospheres. The dynamics of the atmospheric gas is driven by the
  irradiating host star and/or by the rotation of the objects that
  changes during its lifetime. Thermal  gas ionisation in these
ultra-cool but dense atmospheres allows electrostatic interactions and
magnetic coupling of a substantial atmosphere volume. Combined with a
strong magnetic field $\gg B_{\rm Earth}$, a chromosphere and aurorae
might form as suggested by radio and X-ray observations of brown
dwarfs. Non-equilibrium processes like cosmic ray ionisation and
discharge processes in clouds will increase the local pool of free
electrons in the gas. Cosmic rays and lighting discharges also
alter the composition of the local atmospheric gas such that tracer
molecules might be identified. Cosmic rays affect the atmosphere
through air showers in a certain volume which was modelled with a 3D
Monte Carlo radiative transfer code to be able to visualise their
spacial extent.   Given a certain degree of thermal ionisation of
  the atmospheric gas, we suggest that electron attachment to charge
  mineral cloud particles is too inefficient to cause an electrostatic
  disruption of the cloud particles. Cloud particles  will therefore
not be destroyed by Coulomb explosion for the local temperature in the
collisional dominated brown dwarf and giant gas planet
atmospheres. However, the cloud particles are destroyed
electrostatically in regions with strong gas ionisation.  The
potential size of such cloud holes would, however, be too small and
might occur too far inside the cloud to mimic the effect of, e.g.,
magnetic field induced star spots.
\end{abstract}

\section{Introduction}

The presence of atmospheric clouds outside the solar system has now
been established through multi-wavelength variability observations for
brown dwarfs and through transit spectroscopy in extrasolar planets as
summarized in \cite{hell2014} and \cite{marley2013}. Theoretical
efforts have focused on modelling the thermodynamic and chemical
structure of such ultra-cool atmosphere in order to predict their
spectral appearance from the optical into the far-infrared spectral
region (\citealt{allard1995,bur1997,marley2002, fort2008, witte2011}).
A keystone in this efforts is the modelling of cloud formation and
cloud feedback on the atmosphere. Clouds impose an opacity that is
considerably larger than that of the gas (main gas opacity sources:
CO/CH$_4$, H$_2$O, TiO/VO, Na, K), and they deplete or enrich the
element abundance inhomogeneously (e.g. Fe, Mg, Si, O, Ti, Al).

Recently, observations in the near infrared (\citealt{sora2014}) and
in H$\alpha$ indicate potential chromospheric emission
(\citealt{schm2015}) which would suggest a outward temperature
increase in the atmosphere of brown dwarf. Previously, brown dwarf
atmospheres were considered too cold to exhibit any characteristic
plasma signature (\citealt{moh2002}). Radio observations of brown
dwarfs support the expectation that the atmospheres of brown dwarfs
exhibit plasma behaviour (e.g. \citealt{halli2015}), and open up the
possibility to study the magnetic behaviour of such ultra-cool
objects. Interpretations of such radio detection at present invoke the
electron cyclotron maser instability as reason for radio emission at 1
- 100GHz which requires a large magnetic field strength at the site
where the cyclotron maser criterion is fulfilled
(\citealt{vorg2011}). As the cyclotron frequency, $\nu_c$, scales with
the local magnetic field strength B as $\nu_c = eB/(2\pi m_e\,c)$, a
magnetic field strength of $>34$kG is required, for example, in the
case of the M9.5 brown dwarf TVLM 513Ð46546 which was observed to emit
at $\sim 100$GHz (\citealt{will2015}) for cyclotron emission to
occur. Typical magnetic field strength for such late M-dwarfs/brown
dwarfs are $O(10^3)$G which is $30\times$ lower than required for
cyclotron emission. A more suitable interpretation seems that the TVLM
513 emission detected by ALAMA originates from a population of near
speed-of-light (weakly relativistic) electrons in the form of
gyrosynchrotron emission, maybe comparable to Auroral km emission on
Uranus. Aurora observations from the solar system start to emerge as
guide for radio emission on brown dwarfs
(\citealt{nic2012,halli2015}). All mechanisms, the cyclotron maser
instability and Auroral emission, however, require a pool of free
electrons that are captured by the magnetic field and therefore can
emit as accelerated charges. The challenge is that no brown dwarf has
been found to orbit a Sun-like host star from which it could pick up
charges from the stellar wind like the solar system planets nor can we
assume without proof that brown dwarfs host geologically active moons
to provide a steady stream of charges like in the case of the Jupiter
moon Io. Studies of possible ionisation processes occurring in the
ultra-cool atmospheres of brown dwarfs are therefore conducted, and
Section~\ref{s:ogre} discusses  processes that effect the
ionisation state of ultra-cool atmospheres.  Main results of a recent
reference study will be summarised against which the effect of
additional processes, like Cosmic Ray ionisation or lightning
discharges in clouds, can be compared for understanding the radio
emission from ultra-cool atmospheres.  The {\sc Drift-Phoenix}
atmosphere models utilised here as input for the local gas
temperature, gas pressure, thermal electron number density and cloud
properties are 1D atmopshere simulations
(\citealt{helling2008,helling2008e,witte2009,witte2011}).  We
therefore performed 3D Monte Carlo radiative transfer simulations of
Cosmic Ray induced air showers to understand how spatially extended
the effect of Cosmic Rays on the local atmospher might be  and how
  this may differ from our previous 1D results. Section~\ref{s:duch}
investigates the charging of mineral cloud particles through thermal
collisions with the atmospheric gas to provide a first insight into
the possible number of charges that a mineral cloud particle can carry
in comparison to their stability against charge induced destruction
processes.  We argue that regions with strong local gas ionisation,
like Alfv{\'e}n ionisation, would lead to the destruction of the cloud
particles in this regions which could then mimic the appearance of
cold spots on the brown dwarf's surface. Electrostatic cloud
destruction would also help to understand the increasing variability
across the spectral L-T transition for brown dwarfs
(e.g. \citealt{gizis2015}). However, our scale estimates show that the
resulting cloud holes are too small and potentially too deep inside
the cloud to have an observational effect.

\section{Sources of free electrons in ultra-cool atmospheres}\label{s:ogre}

\begin{figure}
\centering
\hspace*{-0.9cm}\includegraphics[angle=0,width=0.6\textwidth]{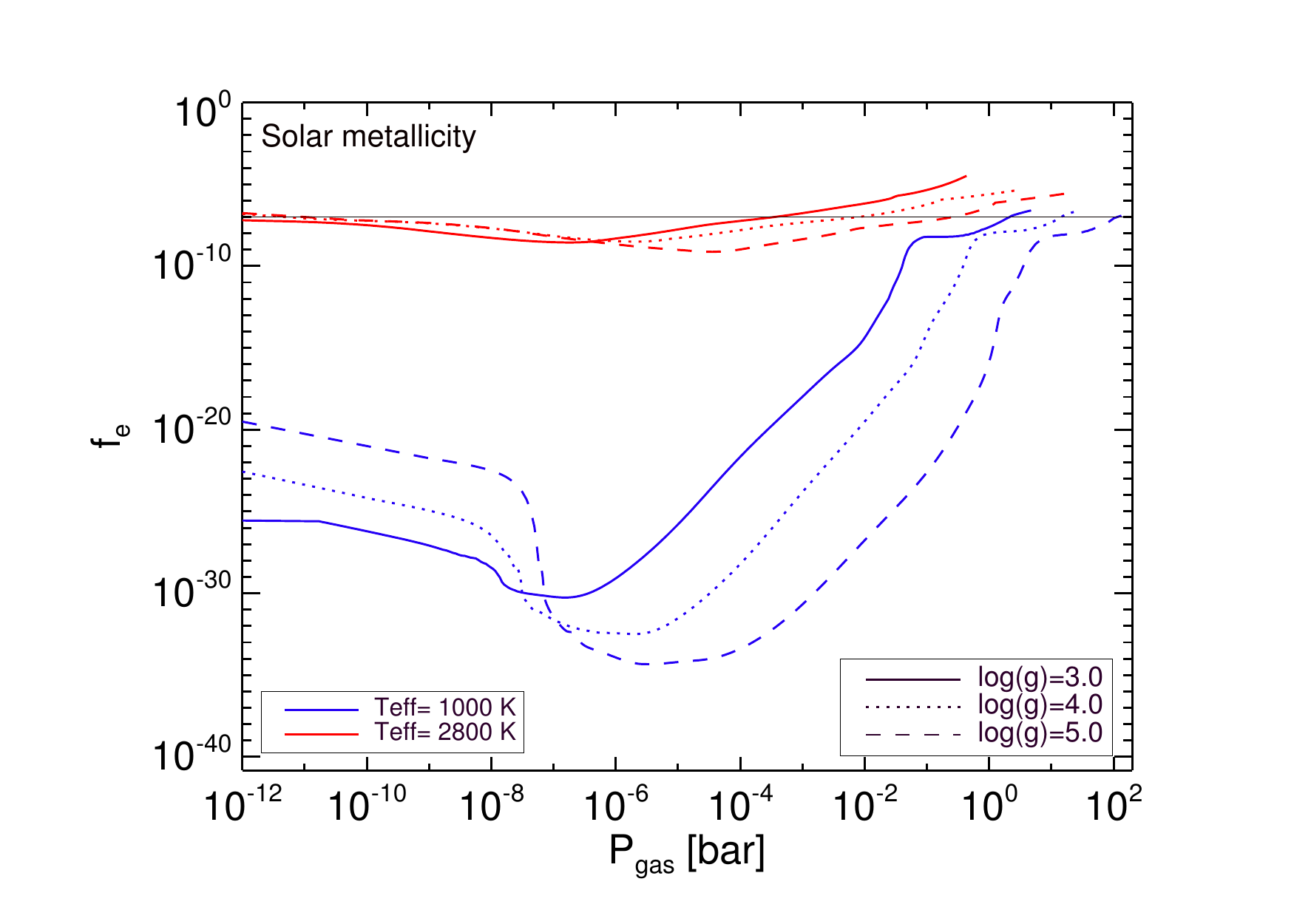}\\*[-0.8cm]
\hspace*{0.2cm}\includegraphics[angle=0,width=0.505\textwidth]{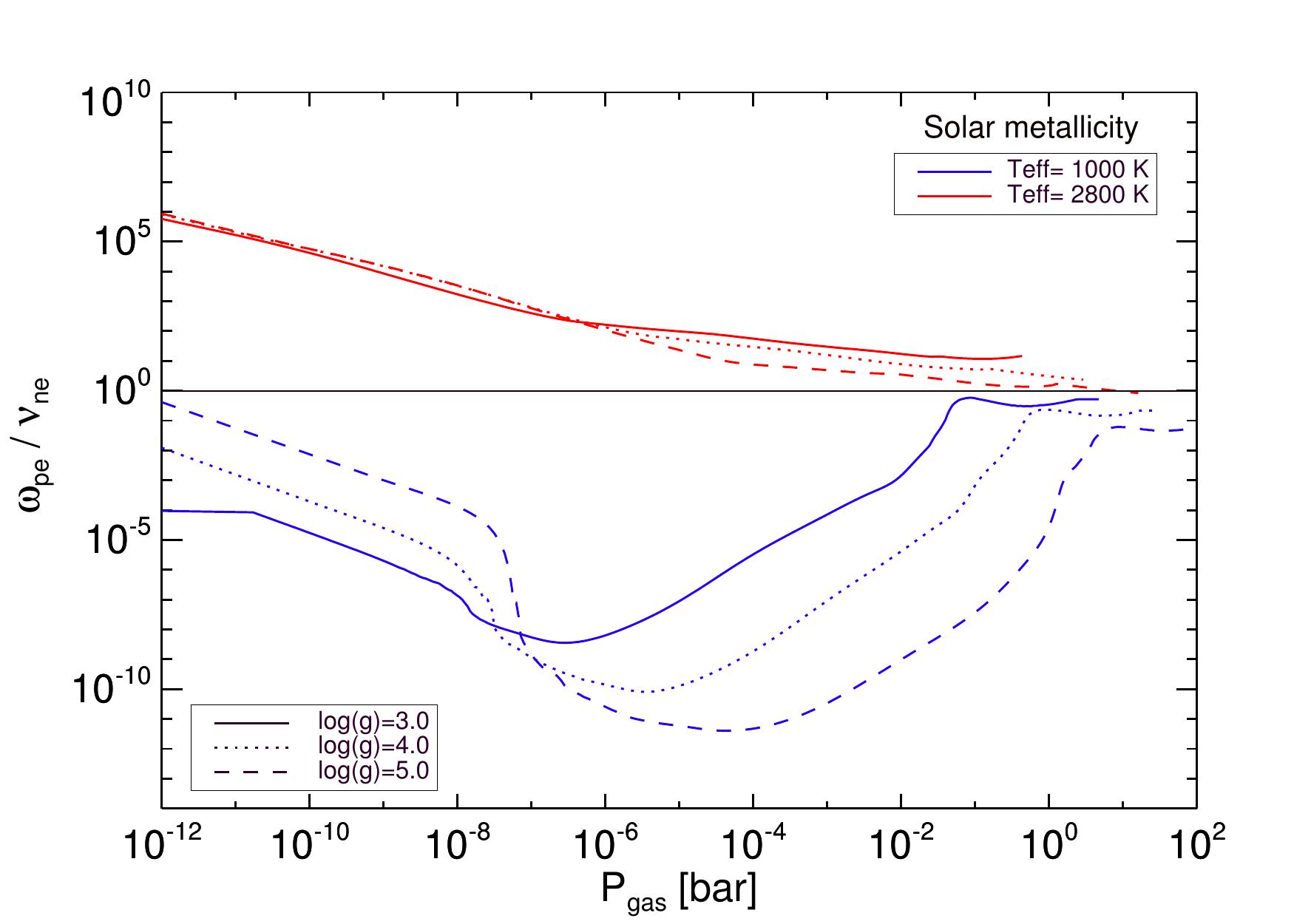}\\*[-0.8cm]
\hspace*{-0.9cm}\includegraphics[angle=0,width=0.6\textwidth]{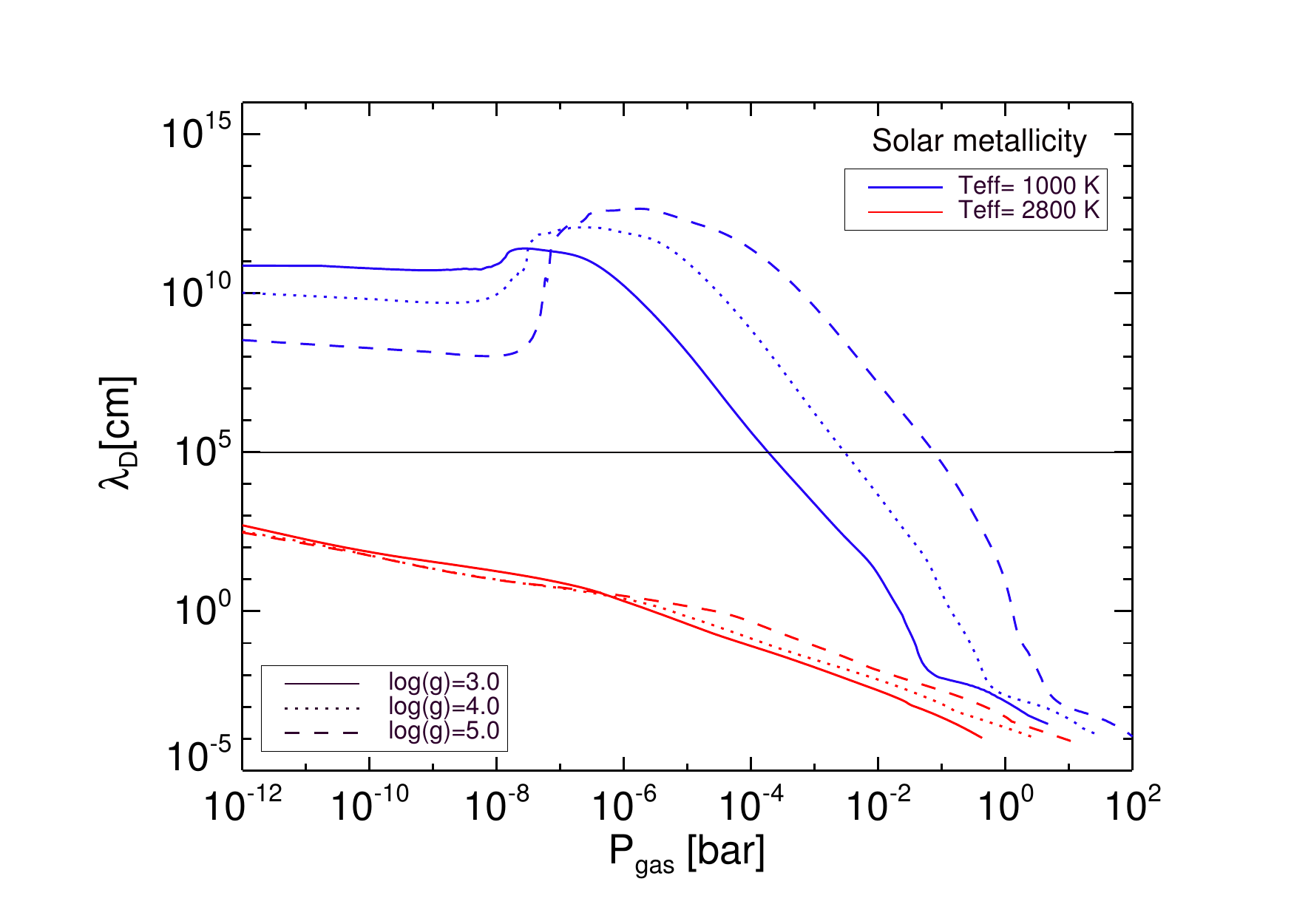}\\*[0.05cm]
\caption{Characterising the thermal ionisation state in ultra-cool atmosphere:
  {\bf Top:} local degree of thermal ionisation $f_{\rm e}$, {\bf Middle:}
  plasma frequency vs. kinetic collisional frequency between electrons
  and neutrals, $\omega_{\rm pe}/\nu_{\rm ne}$ ($\nu_{\rm
    ne}=\sigma_{\rm gas}n_{\rm gas}v_{\rm th, e}$, $\sigma_{\rm gas}$
  -- scattering cross section, $v_{\rm th, e}=\sqrt{k_{\rm B}T_{\rm
      e}/m_{\rm e}}$ -- thermal velocity of electrons, $k_{\rm B}$ --
  Boltzmann constant)}
 \label{fig:logg}
\end{figure}

\subsection{Thermal ionisation}\label{ss:ti}  

Thermal ionisation provides the background ionisation in every
atmospheric gas according to the local gas temperature and pressure.
A reference study of the ionisation and magnetic coupling behaviour
for the atmospheres of ultra-cool objects considering thermally
ionisation can be conducted on the base of existing model atmosphere
grids. We utilise a grid of {\sc Drift-Phoenix} model atmosphere
simulations which include cloud formation modelling. The grid models
cloud forming atmospheres of M dwarfs, brown dwarfs and giant gas
planets. \cite{rod2015} use the whole set of models, here we only
apply a subset. Each atmosphere model is determined by global
parameters that unambiguously characterise a specific object. The
global parameter for a giant gas planet atmosphere model would be the
total flux T$_{\rm eff}=$1000K, the surface gravity log(g)=3.0; for a
brown dwarf it would be T$_{\rm eff}=$1000K, the surface gravity
log(g)=5.0, plus a set of element abundances. A M dwarf (or a young
brown dwarf) would have a higher effective temperature, T$_{\rm
  eff}=2800$K, and a lower surface gravity of log(g)=4.0(\footnote{see
  https://leap2010blog.wordpress.com/category/drift-phoenix/ for a
  summary of {\sc Drift-Phoenix} atmosphere modelling}). The (initial)
element abundances are assumed to be solar but they are altered by
element depletion or enrichment because of cloud formation or
evaporation. \textsc{Drift-Phoenix} solves equations for
non-equilibrium kinetic cloud formation (\citealt{woi2003,helling2006,
  helling2008a,hell2013a}), hydrostatic and chemical equilibrium, and
uses mixing length theory and radiative transfer theory to calculate
the temperature structure (\citealt{hau1999}).
\textsc{Drift-Phonenix} describes the formation of cloud particles as
a phase transition process by considering seed formation, grain growth
and evaporation, sedimentation, element depletion and the feedback of
these processes on the atmosphere structure
(\citealt{woi2004,helling2008,helling2008e, witte2009,witte2011}). The
resulting 1D atmosphere structures are characterised by their local
gas temperature, $T_{\rm gas}$ [K], and gas pressure, $p_{\rm gas}$
[bar], but also by a cloud structure with results for material
composition of the cloud particles, cloud particle size, cloud
particle and gas-phase number densities, all depending on
height. These values allow to derive the local degree of ionisation
$f_{\rm e}(z)=n_{\rm e}(z)/n_{\rm gas}(z)$ ($n_{\rm gas}(z)=p_{\rm
  gas}(z)/(k_{\rm B}T_{\rm gas}(z))$ -- total gas number density
[cm$^{-3}$], $k_{\rm B}$ -- Boltzmann constant, $n_{\rm e}(z)$ --
electron number density [cm$^{-3}$]), the plasma frequency,
$\omega_{\rm pe}=\sqrt{n_{\rm e}e/(\epsilon_0m_{\rm e})}$ ($e$ --
electron charge [C], $m_{\rm e}$ -- electron mass), and the Debye
length $\lambda_{\rm D}=\sqrt{\epsilon_0k_{\rm B}T_{\rm e}/(n_{\rm
    e}e^2)}$.  A certain ionisation is required for a plasma to
establish long-distance electrostatic
interactions. Figure~\ref{fig:logg} (top) demonstrates that the local
degree of thermal ionisation is rather low throughout the
low-temperature atmospheres depicted. It, however, reaches a certain
threshold of $f_{\rm e}>10^{-7}$ (horizontal line) in the inner (high
pressure) atmosphere above which experiments suggest plasma behaviour
to set in. The comparison between the plasma frequency, $\omega_{\rm
  pe}$, and the electron-neutral collision frequency, $\nu_{\rm ne}$,
demonstrate that the upper, low pressure part of the atmosphere is
most susceptible to long-range electrostatic interactions, i.e. where
$\omega_{\rm pe}/\nu_{\rm ne}\gg1$. The Debye length shows over which
length scale such electrostatic interactions can be expected to occur
and Fig.~\ref{fig:logg} (bottom) shows that they are particularly
large for the atmospheres with the lowest effective temperature
T$_{\rm eff}$, in the upper atmosphere where the degree of ionisation
is smallest.

\begin{figure}
\centering
\hspace*{-0.9cm}\includegraphics[angle=0,width=0.55\textwidth]{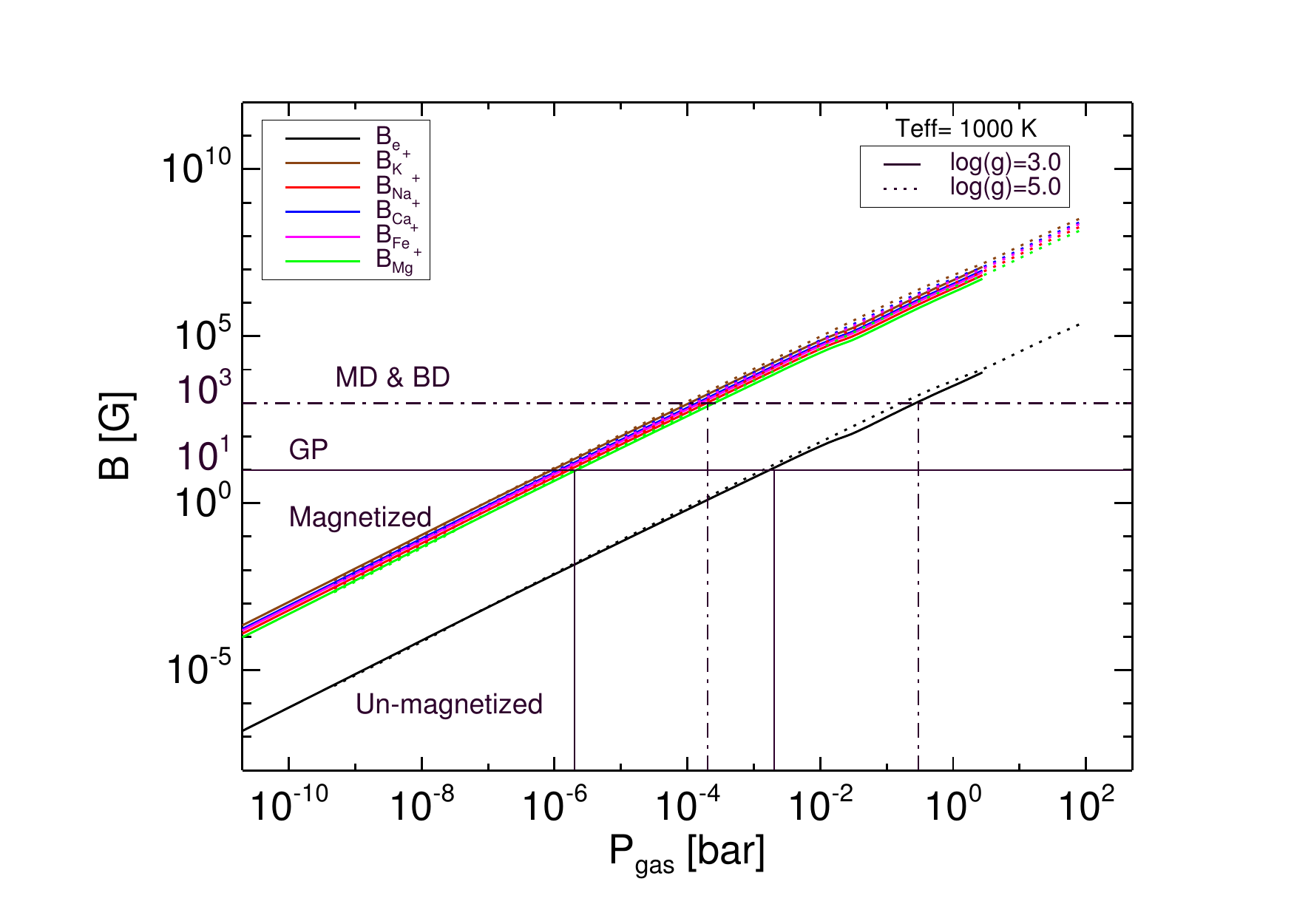}
\caption{The magnetic coupling of ultra-cool and thermally
    ionisaed atmospheres. Horizontal lines indicate typical magnetic
  field densities for M dwarfs and Brown Dwarfs ($10^3$G), and giant
  gas planets (10\,G). Vertical lines link these values with the
    gas pressure where these field strength would occur inside the
    atmosphere. The equivalent global Earth magnetic field strengths
  is $\approx 0.3$\,G.}
 \label{fig:Bcrit}
\end{figure}

For a charged particle's motion to be dictated by a magnetic field,
and hence producing radio emission, the particle needs to complete a
considerable number of gyrations before a collision with a neutral
atom occurs. If this charged particle is, for example, an
  electron, the comparison between the electron cyclotron frequency,
$\omega_{\rm c, e}=eB/m_{\rm e}$ and the electron-neutral collision
frequency, $\nu_{\rm n,e}$ allows to derive a critical value for the
local magnetic field for a magnetic coupling of the gyrating ionised
species. The critical local magnetic field for electrons to be
magnetically bound is $B_{\rm e}\gg (m_{\rm e}/e)\sigma_{\rm
  gas}n_{\rm gas}\sqrt{k_{\rm B}T_{\rm e}/m_{\rm e}}$ which is
independent on the local state of gas ionisation (for more details see
\citealt{rod2015}). Figure~\ref{fig:Bcrit} shows that the upper parts
of ultra-cool atmospheres are magnetically coupled for both,
  electrons and atomic ions, in brown dwarfs and giant gas planets. A
comparison to the magnetic field densities expected for M dwarfs,
brown dwarfs and giant gas planets (horizontal lines) shows that
a larger atmospheric volume can be magnetically coupled in a brown
dwarf atmosphere than in a giant gas planet atmosphere.

\subsection{Non-thermal processes: winds, clouds and cosmic rays}\label{ss:nt}

Brown dwarfs are fast rotators which causes the atmosphere to develop
winds. Giant gas planets develop strong winds due to the strong
irradiation by their host stars. Global circulation models for giant
gas planets (\citealt{show2008, dobbs2010, dobbs2013, show2015})
suggest local wind speeds of several km s$^{-1}$. High wind speeds are
reached in the equatorial jet streams in the upper atmosphere of the
giant gas planet HD\,189\,733b, for example. \cite{zhang2014}
demonstrate, however, that the global wind speed does not exceed
0.2-0.5 km s$^{-1}$ in brown dwarf atmospheres based on their present
set of 3D atmosphere simulations.  \cite{diver2005} demonstrate that a
wind speed of $2-5$ km s$^{-1}$ is required to allow the
collisions of the wind with the ions to cause a charge imbalance
resulting in a considerable increase of the local degree of
ionisation. The required magnetised seed plasma will keep the
electrons locked in place by a magnetic field to allow the wind to
push away the ions which have a considerably larger collisional cross
section than electrons. Therefore, the local charge imbalance imposed
by the gas flow must be established on a timescale, $\tau_{\rm
  s}=m_{\rm i}/(q_{\rm i}B)$ ($\omega_{\rm i}=q_{\rm i}B/m_{\rm i}$ --
ion cyclotron frequency, $q_{\rm i}$ -- ion charge, $m_{\rm i}$ -- ion
mass), shorter than that for electron transport to neutralise it
again.  \cite{stark2013} apply this idea to brown dwarf atmospheres.
However, this time scale might be more appropriately represented by
the collisional frequency with the neutral gas as brown dwarf/ giant
gas planet atmospheres have very high densities, $\tau_{\rm s} \sim
\nu_{\rm ni}$ with $\nu_{\rm ni} \approx \pi r_{\rm H_2}^2 \times
n_{\rm gas} v_{\rm th, i}$, $v_{\rm th, i} = (k_{\rm B}T_{\rm
  i}/m_{\rm i})^{-1/2}$, and $T_{\rm i}=T_{\rm gas}$. Assuming that
the ions follow the wind, the size of the pocket of such an Alfv{\'e}n ionised gas can be
estimated from $R_{\rm Alf} \approx v_{\rm wind}/\nu_{\rm ni}$.

\begin{figure*}
\centering
\hspace*{-0.9cm}\includegraphics[angle=0,width=0.5\textwidth]{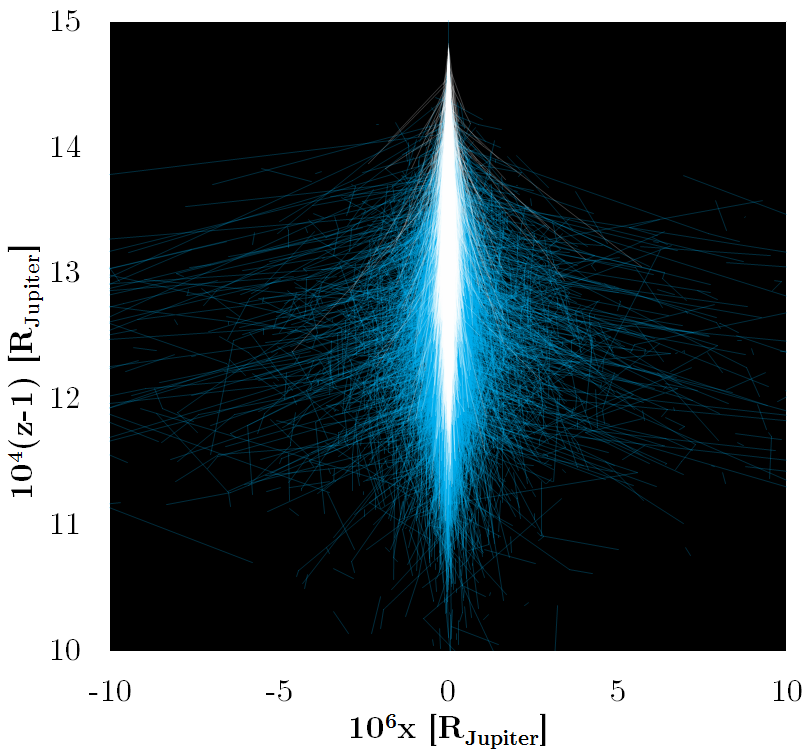}
\hspace*{-0.9cm}\includegraphics[angle=0,width=0.5\textwidth]{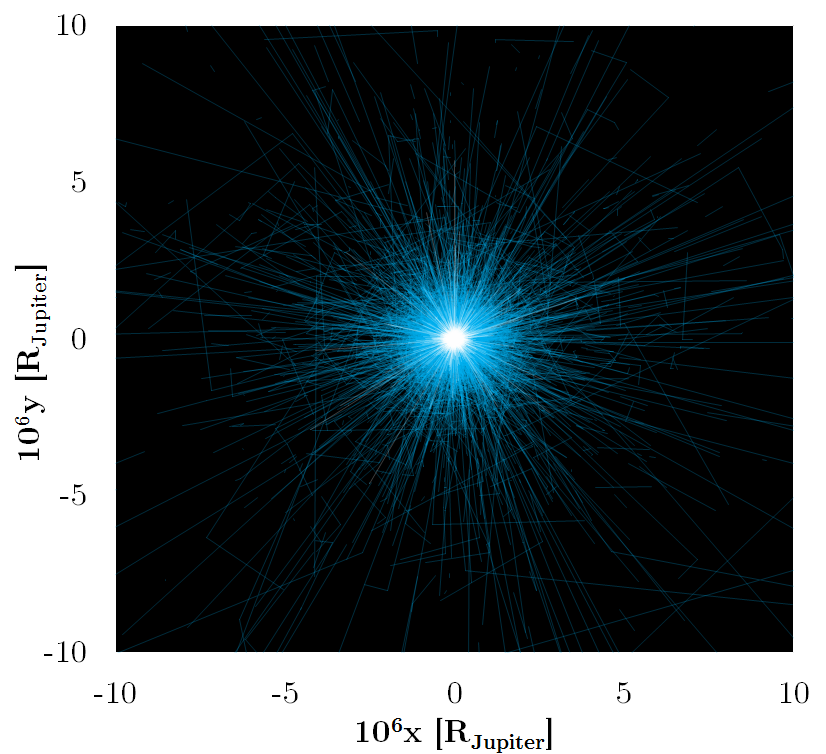}
\caption{3D Monte Carlo simulation of a Cosmic Ray triggered air
  shower in a Jupiter-like atmosphere. The proton trajectoriers (blue)
  and the positive pions (white) are triggered by a primary CR event
  of $10^{20}$eV. {\bf Left:} z-y plane view, {\bf Right:} view in
  direction of motion with the CR at an zero angle into the atmosphere
  (x-y plane). The figure demonstrates that the air shower
    remains rather confined with respect of the atmospheric volume
    effected. The use of 1D simulation to calculate the CR spectrum impact on
    the electron budget appear therefore suitable.}
 \label{fig:airshower}
\end{figure*}
 
Figure~\ref{fig:Bcrit} shows that in particular the upper,
low-pressure regions of an atmosphere can be expected to be
magnetically coupled. Alfv{\'e}n ionisation will therefore work best
at gas pressures $<1$bar in brown dwarf atmospheres and
$<10^{-2.5}$bar in giant gas planet atmospheres. This pressure range
coincides with the atmospheric range where mineral clouds form in
brown dwarfs and giant gas planets (Fig.~2 in
\citealt{hell2008}). This suggests that the charge imbalance caused by
the impact of strong winds on a magnetised plasma could contributes to cloud particle ionisation in these pockets of Alfv{\'e}n ionisation.

\cite{hell2011a} argue that clouds in brown dwarf and giant gas planet
atmosphere are electrostatically charged because particle-particle
collisions in turbulent atmosphere clouds alone are energetic enough
to overcome the work function of mineral materials (triboelectric
charging). The size-dependent gravitational settling (rain-out), which
determines the cloud height, causes a large scale charge separation
resulting in an electrostatic potential difference inside such a
mineral cloud. \cite{hell2013} demonstrated that such a potential
difference can overcome the breakdown field and, hence, initiate
an ionisation avalanche that subsequently can lead to a large-scale
lighting discharge inside mineral clouds. \cite{bail2014} use scaling
laws derived from sprite experiments and demonstrate that such
large-scale discharges can reach an geometrical extension of $\approx
3000$km in brown dwarf clouds but only $\approx 300$km in a giant gas
planet atmosphere.

Cosmic rays (CR) contribute to the ionisation of clouds on Earth and
they are discussed as trigger for lighting initiation in the Earth
atmosphere (\citealt{Gurevich2013}). Similar effects occur in brown
dwarfs and extrasolar planets (\citealt{rim2013}) where each of the
systems may be exposed to a different radiation field of a host star
or the interstellar medium irradiated by a high-mass O- or B-type
star. For the interstellar cosmic ray flux spectrum (e.g., Fig. 4 in
\citealt{rim2013}), the local degree of gas ionisation increases by
$\approx$ 6 orders of magnitude in the case of a giant gas planet
atmosphere and by $\approx$ 4 orders in a brown dwarf atmosphere that
is $10^2\times$ more compact. Cosmic rays do also effect the abundance
of molecules through ion-neutral kinetic gas chemistry by opening up
reaction channels to more and more complex hydro-carbon molecules
(\citealt{rim2014IJA}). The investigation of the spatial extent of CR
triggered events is required to help determining how much of the
atmosphere volume might produce chemical tracers. So far, only 1D
  simulations were done for extrasolar objects  (\citealt{rim2013}).

\begin{figure}
\centering
\hspace*{-0.0cm}\includegraphics[angle=0,width=0.45\textwidth]{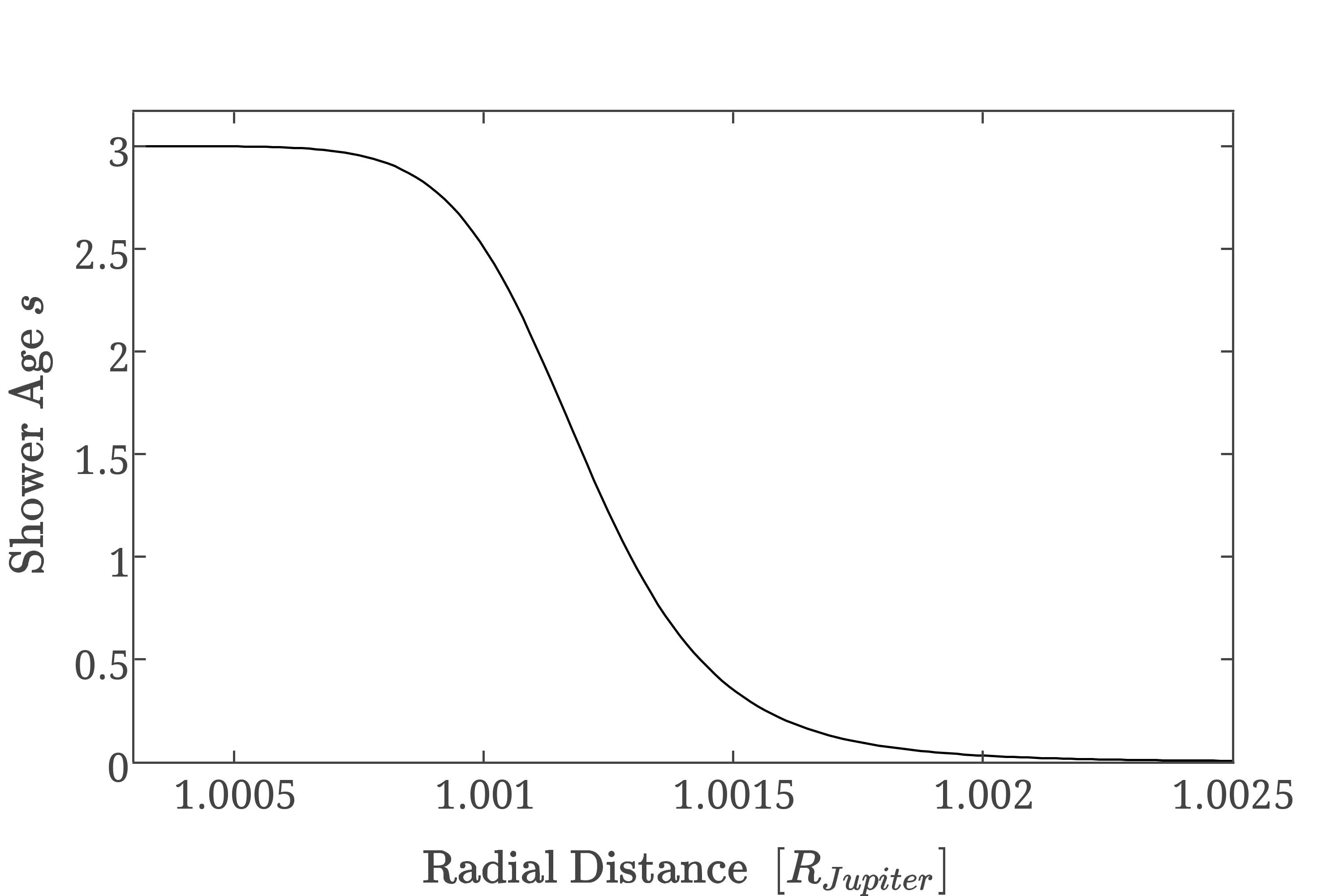}
\caption{The air shower penetration depth (shower age), $s=s(X)$ (X - column density) for a Jupiter-like atmosphere and a CR event of $10^{20}$eV:
  $s=0$ corresponds to the initial event, $s=1$ corresponds to the
  shower maximum occuring at r=1.0013 R$_{\rm Jupiter}$, and $s=3$ for
  in infinit column density $X\rightarrow\infty$.}
 \label{fig:s}
\end{figure}

We therefore carried out first 3D radiative transfer calculations
utilizing the Monte Carlo method (\citealt{wood1999}) which has been
updated to include algorithms for particle tracking from
\cite{dupree2002} as a follow-up of the work presented in
\cite{rim2013} for hydrogen-rich atmospheres. We model the
  interactions of protons, positive, neutral and negative pions,
  positive and negative muons, electrons and positrons, and gamma rays
  with an H$_2$-dominated gas in a three dimensional Cartesian grid
  representing a homogeneous atmosphere with a radial
  density-temperature gradient.  The atmosphere profile that
  determins the density-temperature gardient is the same like in
\cite{rim2013} (their Fig. 3) and is used as input for the local
gas number density.  The 3D atmosphere is assumed to be isotropic, the
density profile only changing with the distance from the center as
prescribed by the 1D profile. The air shower is initiated by one
  CR event which is assumed to have an energy of $10^{20}$eV.  \cite{rim2013}
  demonstrated that high-energy CR events penerate deeper into the
  atmosphere. This approach therefore allows us to study the maximum
  effect on the atmospheric volume influenced by one air shower
  event. \cite{rim2013} have studied the CR electron production rate
  for a whole spectrum of intial CR energies but in 1D. Here, we are
  interested in the geometrical extent of such an event skimming the
  atmosphere for allowing a maximum of observable volume. For an air
  shower intiated by one CR event, the number of particles exceeds a
  million within just a few generations after initiation. We apply a
  thin sampling method where all particles below a predefiened
  thinning energy level are subject to thin sampling with a chance of
  survival proportional to their energy (Hillas 1985; Hillas
  1997). The shower development can then be studied, for example, by
  measuring the penetrative depth of an air shower, called shower age
  $s$, which depends in the integrated column density (see Eq. 3 in
  \citealt{rim2014}). The shower age is, for example, used to
  parameterise the energy distribution of secondary electrons (see
  Eq. 4 in \citealt{rim2014}). The shower age can only be
  determined in one direction, and Figure~\ref{fig:s} shows the shower
  age for the radial direction. The air shower reaches its maximum at
  a radial depth of r=1.0013 R$_{\rm Jupiter}$ for a Jupiter-like
  atmosphere. This value is expected to be lower for denser
  atmospheres like for brown dwarfs, hence the impact of CRs on the
  electron gas density will be smaller in Brown Dwarfs than in
  Jupiter-like planets.

Figure~\ref{fig:airshower} shows results in 2D cuts for an air shower
triggered by an CR event of $10^{20}$eV. Note that the longitudinal
extension of the shower is approximately two orders of magnitude
greater than the lateral extension.  The opening angle of the air
shower remains rather confined despite various singular trajectories
branching out.  The shower has no significant anisotropies in the x-y
plane (Fig.~\ref{fig:airshower}, right). Images of the other particle
types appear vastly different, for example, the trajectories of
neutral pions are mostly sub-centimetre because they decay almost
instantaneously, hence, only effect the uppermost atmosphere.

Our 3D results confirm that also in extrasolar objects only the
uppermost atmospheric layers will be affected by cosmic air showers
because of the rather high local densities in the atmospheres of giant
gas planets and brown dwarfs. Their occurrence statistics will depend
on the external cosmic ray flux which will be higher for young brown
dwarfs in a star forming regions with strongly radiating O- and
B-stars or for an object near a super nova outburst compared to an
extrasolar planet in a planetary system around a Sun-like star.

%{\bf summerise everything into one picture}

\section{Charging  cloud particles by atmospheric gas interaction}\label{s:duch}

Atmospheres of extrasolar planets have been shown to form clouds that
are made of mixed mineral particles of various sizes (e.g.,
\citealt{hell2008}). The growth of these cloud particles is determined
by collisions with the ambient gas which they deplete. Cloud particles
start to gravitational settle and change their size and composition
depending on the local thermodynamic conditions. The distance over
which the cloud particle ensemble precipitates does determine the
geometrical height of the cloud.  Cloud particle will not only collide
with neutral species, but also with electrons and ions and by this
process pick up charges from the surrounding atmospheric gas. The
efficiency of this charging process will depend on the abundance of
the electrons and ions in the gas phase and their temperature. In
extrasolar, ultra-cool atmosphere, the amount of free charges will be
moderate as discussed in Sect.~\ref{ss:ti} if thermal ionisation is
the only gas ionisation process. Section~\ref{ss:nt} summarised other
processes that act to increase the pool of free charges in ultra-cool
atmospheres. A background of thermal electrons will therefore always
be present. In following section, we offer a first exploration of how
such atmospheric charges may effect the cloud particle charges, and
argue that regions of strong gas ionisation destroy cloud particles
electrostatically. If such a destruction occurs, the cloud
  opacity would change and possibly allow for an observable change of
  radiation flux.

\subsection{Electron Deposition on cloud particles in ultra-cool atmospheres} \label{model}
We perform a first exploratory estimate of atmospheric regions of
extrasolar planets and brown dwarfs where cloud particles are most
likely to be stable against electrostatic disruption caused by charges
accumulation on their surfaces.  As a first attempt to evaluate cloud
particle charging in brown dwarf atmospheres, electron deposition on
cloud particle is considered as a first order process where the
respective rate coefficient depends only linearly on the gas density,
$k\sim n_{\rm gas}$. Gas-phase ionisation mechanisms were
  discussed in Sect~\ref{s:ogre} for brown dwarfs and extrasolar
  planetary atmospheres. The cloud particles are assumed to be
spherical. The gas-phase ionisation is parameterised by an ionisation
rate $\zeta$ [s$^{-1}$].  Electrons are adsorbed onto the surface of a
cloud particle with a certain probability when the electron in the
atmospheric gas collides with the cloud particle.
%, unless they recombine with a cation before they can
%reach the cloud particle surface. 
Similar to the growth of cloud particles by gas-surface reactions
(\citealt{woi2003}), the electron will approach the surface directly
(free molecular flow) or diffusive (viscous) before the electrostatic
interaction dominates. We assume that electrostatic
attraction/repulsion will dominate the approach of the electron
or ion. Considerably more elaborate approaches have been published for
dust charging in protoplanetary disks by
e.g. \cite{fujii2011,ilg2012}.  We aim to provide this first estimate
in combination with our detailed modelling of cloud formation which
provides the local cloud particle number density, $n_{\rm d}$
[cm$^{-3}$], their size, $a$ [cm], and material composition which is a
mix of materials (\citealt{woi2004,helling2006,helling2008a}), in
contrast to previous works. Mineral clouds in brown dwarfs and
  giant gas planets are characterised by cloud particles changing in
  size and material composition depending in the local temperature and
  gas density. Gnerally, the top of the cloud is dominated by small
  particles ($10^{-6}$cm) and a mix of Mg/Si/Fe/O materials (see Tabe
  7 in \citealt{hell2008}). The cloud base can be made of particles a
  large as $10^{-2}$cm which predominently contain high-temperature
  condensates made of Fe/Al/Ti/O. These findings are based on a
  kinetic cloud formation models that described seed formation and
  subsequent surface growth/evaporation by gas phase - surface
  reactions. The model derived by
  \cite{woi2004,helling2006,helling2008a} does fulfil element
  conservation and takes into account gravitational settling and
  convective mixing for the cloud particle formation.

In the following, a locally constant grain size is assumed. Changes in
grain size could inprinciple occur but are not considered in our
following explorative estimate.

The net number of electrons per grain (net negative charge), $N_{\rm
  e,d}$, changes depending on the local electron number density,
$n_e$, through electron and ion adsorption as
\begin{equation}
\dfrac{dN_{\rm e,d}}{dt} = (k_- - k_+)n_e.
\label{eq:surface-rate-eq}
\end{equation} 
The electron number density, $n_{\rm e}$,  in the gas phase  changes therefore through cloud particle ionisation as
\begin{equation}
\dfrac{dn_e}{dt} = \zeta n_{\rm gas}  - \alpha n_e^2 - k_-n_{\rm d}n_e.
\label{eq:gasphase-rate-eq}
\end{equation}
The first term parameterises the gas ionisation by a prescribed
ionisation rate $\zeta$ [s$^{-1}$]. The second term describes the
gas-phase recombination the third term is the loss through
the adsorption of an electron onto the grain surface. The total rate
coefficient for recombination in the gas-phase, $\alpha$ [cm$^3$
  s$^{-1}$], can be written as
\begin{equation}
\alpha = \alpha_2 + n_{\rm gas} \alpha_3,
\label{eq:diss-recom}
\end{equation}
where $\alpha_2$ [cm$^3$ s$^{-1}$] is the 2-body gas-phase recombination rate, and $\alpha_3$ [cm$^6$ s$^{-1}$] is the 3-body rate:
\begin{align}
\alpha_2 &= 8.22 \times 10^{-8}\Bigg(\dfrac{T}{\rm 300K}\Bigg)^{\!\!-0.48} - 1.3\times10^{-8},
\label{eq:2-body}\\
\alpha_3 &= 2 \times 10^{-25}\Bigg(\dfrac{T}{\rm 300K}\Bigg)^{\!\!-2.5}.\label{eq:3-body}
\end{align}
The adsorption rate for electrons onto a cloud particle, $k_-$, and
for ions $k_+$, both with units [cm$^3$ s$^{-1}$], can be expressed as
\begin{align}
k_- &= \sigma_{\rm gr}\sqrt{v_e^2 - \dfrac{N_{\rm e,d}e^2}{2\pi \epsilon_0 m_e a}}, \label{eq:e-dep}\\
k_+ &= \sigma_{\rm gr}\sqrt{\dfrac{8k_{\rm B}T}{\pi m_p} + \dfrac{N_{\rm e,d}e^2}{2\pi \epsilon_0 m_p a}},  \label{eq:ion-dep}
\end{align}
where $k_{\rm B} = 1.38 \times 10^{-16}$ erg/K is Boltzmann's
constant, $v_e$ [cm/s] is the electron velocity, $m_e = 9.11 \times
10^{-28}$ g, and $m_p = 1.67 \times 10^{-24}$ g. The cross-section for
a spherical dust grain is $\sigma_{\rm gr} = \pi a^2$, with $e$ the
elementary charge and $\epsilon_0$ the permittivity of free space. The
second terms in Eqs.~\ref{eq:e-dep},~\ref{eq:ion-dep} account for the
effect of the electrostatic field of the negatively charged grain upon
the gas-phase electrons and ions.  The free electrons are not
necessarily in thermal equilibrium with the ambient neutral gas, or
the ions. Brown dwarf atmospheres are, however, very dense and
therefore collisional dominated. We use $T_{\rm e}=T_{\rm i}=T_{\rm
  gas}$ unless stated otherwise. {The velocity of the electron is
  therefore taken to be its thermal velocity, $v_{\rm e}=v_{\rm
    th,e}$.
%$v_e = (8k_BT_{\rm   gas}/\pi m_e)^{1/2}$, and for the ions $v_i = (8k_BT_{\rm gas}/\pi
%m_i)^{1/2}$. 
This will, however, not be true during the development of
a discharge or other high-energy events.} 
% For secondary electrons from
%cosmic ray events, for example, the electron thermal velocity can be
%of the order of $\approx 10^8 - 10^{10}$ cm/s.
%The net negative charge of the dust grain, $N_{\rm e,d}$,
%inhibits the attachment of further electrons and encourages the
%capture of positive ions from the surrounding gas. For a given thermal
%energy, the electrons are more mobile than the ions. Electrons
%therefore have a greater likelihood of interacting with the cloud
%particle surface.  
As the number of surface electrons grows, the
resulting electrostatic field of the cloud particle begins to inhibit
further electron attachment. However, the field attracts ions and
now the probability of ion attachment increases.  Ion attachment
neutralises the grain, lowering the net negative charge of the grain.

{\it The steady-state solution} of Eq.~(\ref{eq:surface-rate-eq}),
$dN_{\rm e,d}/dt = 0$, is satisfied when $k_- = k_+$
(Eqs. \ref{eq:e-dep}, \ref{eq:ion-dep}). The maximum value, $N_{\rm
  e,d, max}$, of charges that a cloud particle of size $a$ can acquire
by adsorption from a surrounding ionised gas of temperature $T=T_{\rm
  gas}$ is therefore
\begin{equation}
N_{\rm e,d, max} = \Big(\dfrac{16 k_B \epsilon_0}{e^2}\Big) aT.
\label{eq:maximum-charge}
\end{equation}
The quantity $N_{\rm e,d, max}$ denotes the number of electrons a
grain will have at steady state. $N_{\rm e,d, max}$ is independent of
the gas ionisation rate $\zeta$ (Eq.~\ref{eq:gasphase-rate-eq}).
Processes that requires a critical surface charge, $N_{\rm d, crit}$,
will occur if $N_{\rm d, crit}< N_{\rm e, d, max}$ (e.g. electrostatic
disruption, \citealt{stark2015}; electron avalanches,
\citealt{hell2011,dub2015}).  If $N_{\rm d, crit}>N_{\rm e, d, max}$,
than the number of electrons on the cloud particle surface would achieve
steady state before the critical process has a chance to occur. The
time needed to achieve steady state is characterised by $k_- \neq
k_+$, and hence $N_{\rm e,d}\ll N_{\rm e,d, max}$. In this case, the
electrostatic contributions to the adsorption rates $k_-$ and $k_+$
become very small, and Eqs.~\ref{eq:e-dep} and~\ref{eq:ion-dep} can be
approximated by
\begin{align}
 k_- &= \sigma_{\rm gr}v_e, \label{eq:e-dep-approx}\\
k_+ &= \sigma_{\rm gr}\sqrt{\dfrac{8kT}{\pi m_p}}. \label{eq:ion-dep-approx}
\end{align}

\subsection{Time evolution of cloud particle changing}\label{solution}
Equation~\ref{eq:gasphase-rate-eq} can be integrated to obtain an
analytic expression for the gas electron number density, $n_e$, that
changes because of electron depletion through grain collisions 
  before steady state is reached or if it is not reached at all
\begin{equation}
n_e = \dfrac{1}{\alpha\tau}\tanh\Bigg[\dfrac{t}{\tau} + \arctanh\Bigg(\dfrac{k_-n_{\rm d}\tau}{2}\Bigg)\Bigg] - \dfrac{k_-n_{\rm d}}{2\alpha},
\label{eq:electron-number}
\end{equation}
with
\begin{equation}
\tau = \dfrac{2}{\sqrt{4\alpha \zeta n_{\rm gas}  + k_-^2n_{\rm d}^2}}.
\label{sol4}
\end{equation}
% overinterpretation
%{\bf the time scale for $N_{\rm e,d}\ll N_{\rm e,d, max}$, i.e. for $dN_{\rm e,d}/dt \not= 0$.}
%wrong:
%the time scale for the cloud particle ensemble of number density
%of $n_{\rm d}$ to reach the charge steady state where $k_- = k_+$.
The first term in Eq.~\ref{sol4} represent the electron-gas
recombination, the second term the electron adsorption onto the
grain's surface.  Inserting Eq.~\ref{eq:electron-number} into
Eq.~\ref{eq:surface-rate-eq} results in
\begin{align}
\dfrac{dN_{\rm e,d}}{dt} & =  \dfrac{k_- - k_+}{\alpha\tau}\tanh\Bigg[\dfrac{t}{\tau} + \arctanh\Bigg(\dfrac{k_-n_{\rm d}\tau}{2}\Bigg)\Bigg] \notag\\
  &  - \dfrac{\big(k_- - k_+\big)k_-n_{\rm d}}{2\alpha}.
\label{eq:surface-explicit}
\end{align}
\noindent
Integrating Eq. (\ref{eq:surface-explicit}) results in an expression
for the total number of charges, $N_{\rm e,d}$, on an ensemble of
cloud particles due to a first-order gas-phase ionization process as a
function of time before steady state is achieved,
\begin{align}
N_{\rm e,d} &= \dfrac{k_- - k_+}{\alpha}\log\Bigg[\cosh\Bigg(\dfrac{t}{\tau} + \arctanh\Big(\dfrac{k_-n_{\rm d}\tau}{2}\Big)\Bigg)\Bigg] \notag\\
& - \dfrac{\big(k_- - k_+\big)k_-n_{\rm d}t}{2\alpha}\notag\\
 &  -\dfrac{k_- - k_+}{\alpha}\log\Bigg[\Bigg(1 - \dfrac{k_-^2n_{\rm d}^2 \tau^2}{4}\Bigg)^{\!\!-1/2}\Bigg].
\label{eq:surface-solution}
\end{align}
 This solution is only accurate when $N_{\rm e,d} \ll N_{\rm e,d,
   max}$, hence before the cloud particles have achieved a charge
 steady state.
% Hence, already when $N_{\rm e,d} \lesssim N_{\rm e,d,
%   max}$, Eq. (\ref{eq:surface-solution}) will overestimate the value
% of $N_{\rm e,d}$. 
The predicted time (Eq.~\ref{sol4}) can only be considered as a lower
limit because the electrostatic effects of the charged grain on free
electrons and cations is neglected in our approximations of
Eq's. (\ref{eq:e-dep-approx})~and~(\ref{eq:ion-dep-approx}) which
  have not been used in the whole of Sect.~\ref{solution}.

For spherical grains the
maximum number of net charges that can reside on a grain of radius $a$
before Coulomb explosion can be expressed as (Eq.~8 in
\citealt{stark2015}),
\begin{equation}
N_{\rm ce}=\dfrac{\pi(32\epsilon_0\Sigma_s)^{1/2}}{e}a^2, \label{coulomb}
\end{equation}
with $\Sigma_{s}$ he mechanical tensile strength of the dust grain
(the maximum stress or pressure that a material material can
withstand). By setting $N_{\rm e,d}=N_{\rm ce}$
(Eqs.~\ref{eq:surface-solution} and \ref{coulomb}), a critical time
scale, $t_{\rm crit}$, can than be evaluated numerically. $t_{\rm
  crit}$ is the time during which the cloud particle can accumulate
charges by electron deposition and still be stable agains Coulomb
disruption. For times larger than $t_{\rm crit}$, i.e.  $t>t_{\rm
  crit}$, cloud particles of a given size will explode due to the
electrostatic force. The solution for $t_{\rm crit}$ is depicted in
Fig~\ref{fig:time-scale}.

\subsection{Grain charge deposition time-scales in ultra-cool atmospheres of extrasolar planets \label{drift}}

{\bf Approach:} We evaluate the maximum charge number densities
possible through charge adsorption and the critical survival time
scale against electrostatic disruption of mineral cloud particles in
the ionised gases of extrasolar atmospheres. We utilised one example
{\sc Drift-Phoenix} atmosphere simulation
(\citealt{witte2009,witte2011}) and use the model results for the
local gas density, $n_{\rm gas}(z)$ [cm$^{-3}$], the gas temperature,
$T_{\rm gas}(z)$ [K] ($\approx 0.1$eV), and the cloud particle size
$a(z)$ [cm] in order to evaluate Eq.~\ref{eq:maximum-charge}, and
the critical time scale,  $t=t_{\rm crit}$, from $N_{\rm e,d}=N_{\rm ce}$ with Eqs.\ref{eq:surface-solution} \&
\ref{coulomb}.  We represent the cloud particle size $a$ by the
height-dependent mean particle size, $\langle a(z)\rangle$ instead of
a height-dependent particle size distribution function. We demonstrate
the results for the model simulation of a giant gas planet atmosphere
with the effective temperature $T_{\rm eff} = 1600$ K (total radiative
flux), the surface gravity $\log g = 3$, and the set of solar element
abundances.  The electron velocity, $v_e(z)$ [cm/s], is calculated
assuming $T_{\rm e}(z)=T_{\rm gas}(z)$ unless stated otherwise. The
first-order ionization rate, $\zeta$ [s$^{-1}$] is used as parameter
and different values are explored.

\begin{figure}
\includegraphics[width=0.5\textwidth]{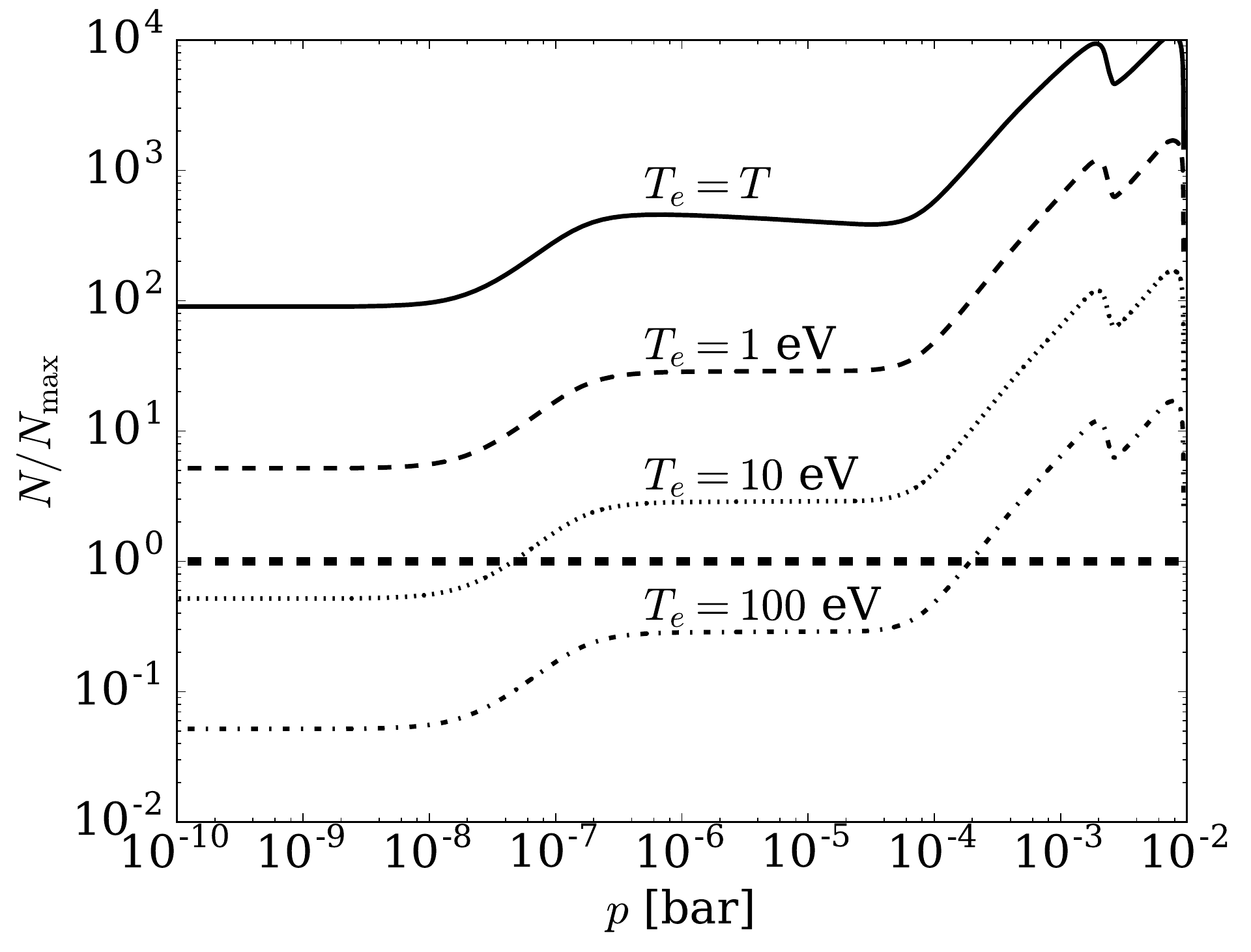}
\caption{The electron number ratio $N/N_{\rm max}$
  (Eq.~\ref{eq:size-limit}) for a mineral cloud formed in a giant gas
  planet or brown dwarf with $T_{\rm eff} = 1600$K, $\log g = 3$ and
  solar element abundance. Different electron temperatures, $T_e =
  T_{\rm gas}$ ($\approx 0.1$eV), 1 eV ($\approx 10^4$K), 10 eV
  ($\approx 10^5$K) and 100 eV ($\approx 10^6$K), are evaluated. The
  horizontal lines represents $N_{\rm ce}/N_{\rm max}=1$. The cloud
  particles are stable against electrostatic disruption if $N/N_{\rm
    max}\gg N_{\rm ce}/N_{\rm max}$, i.e. above the horizontal line.}
\label{fig:size}
\end{figure}

\begin{figure}
\includegraphics[width=0.5\textwidth]{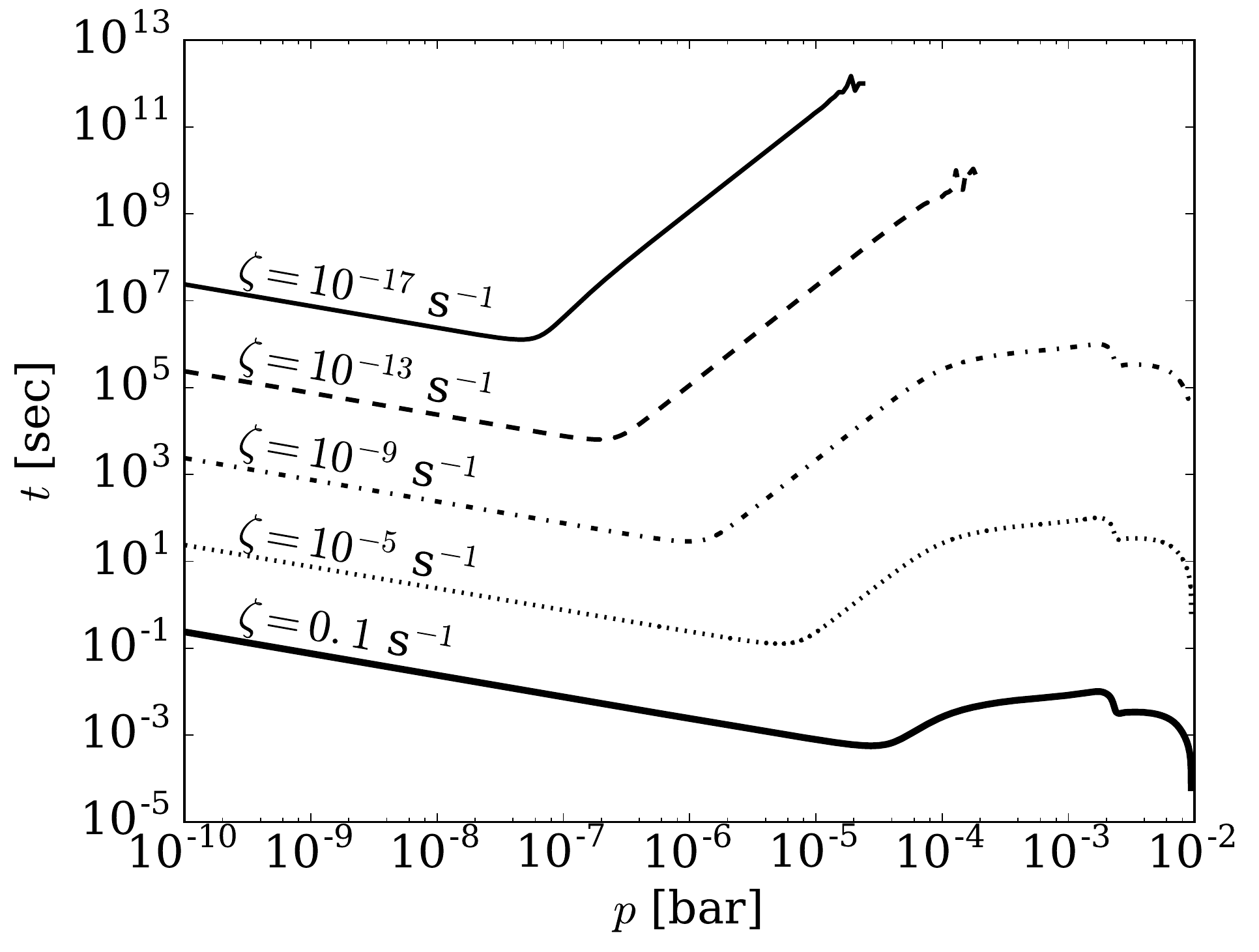}
\caption{The timescale to coulomb explosion, $t=t_{\rm crit}$ [s], 
%for  cloud particles of height-dependent mean size 
 as function of \sed{gas} pressure,
  $p$ [dyn/cm$^2$], for various values of $\zeta$ [s$^{-1}$]. The
  plots result from applying the parameters of a
  \textsc{Drift-Phoenix} model atmosphere ($\log g = 3$, $T_{\rm eff} =
  1600$). Example values are: Cosmic-ray ionization ionization rate:
  $\zeta = 10^{-13}$ s$^{-1}$, extensive air showers: $\zeta \sim
  10^{-9}$s$^{-1}$.}
\label{fig:time-scale}
\end{figure}

{\bf Results:} We first explore the atmospheric regions in our example
giant gas planet/brown dwarf atmosphere where the cloud particles are
stable against electrostatic disruption cause by the charges
accumulated on their surfaces. This can be accomplished by comparing
the number of charges necessary for Coulomb explosion, $N_{\rm ce}$
(Eq.~(\ref{coulomb}), to the maximum possible number of charges that a
cloud particle can adsorb from the ionised atmosphere, $N_{\rm max}$
(Eq.~\ref{eq:maximum-charge}),
\begin{equation}
 \frac{N_{\rm ce}}{N_{\rm max}}= \sqrt{\dfrac{e^2 \Sigma_s}{8 \epsilon_0}}\dfrac{\pi \langle a\rangle}{T_e}
\label{eq:size-limit}
\end{equation}
where $T_e$ [eV] is the electron temperature. Only if $N_{\rm
  ce}/N_{\rm max}\gg1$, the cloud particles are stable against
electrostatic disruption for a given electron temperature, $T_{\rm
  e}$. Coulomb explosion will only take place if $N_{\rm ce}/N_{\rm
  max}\ll1$. Figure~\ref{fig:size} demonstrates that in the model
atmosphere considered here, $T_e$ needs to be much greater than the
thermal energy to achieve $N_{\rm ce}/N_{\rm max}\ll1$. Since the
average composition of the dust will change as a function of
atmospheric height, this value of $\Sigma_s$ will vary between 1 MPa
and 100 MPa, and we set $\Sigma_s = 50$ MPa for the exploratory
purpose of this paper. Generally, the cloud layer is stable against
Coulomb explosion for $T_{\rm e}<10^5$ K ($<10$eV) and at high pressures inside the atmosphere also above $10^5$K (Fig.~\ref{fig:size}).

\begin{figure}
\centering
\hspace*{-0.9cm}\includegraphics[angle=0,width=0.55\textwidth]{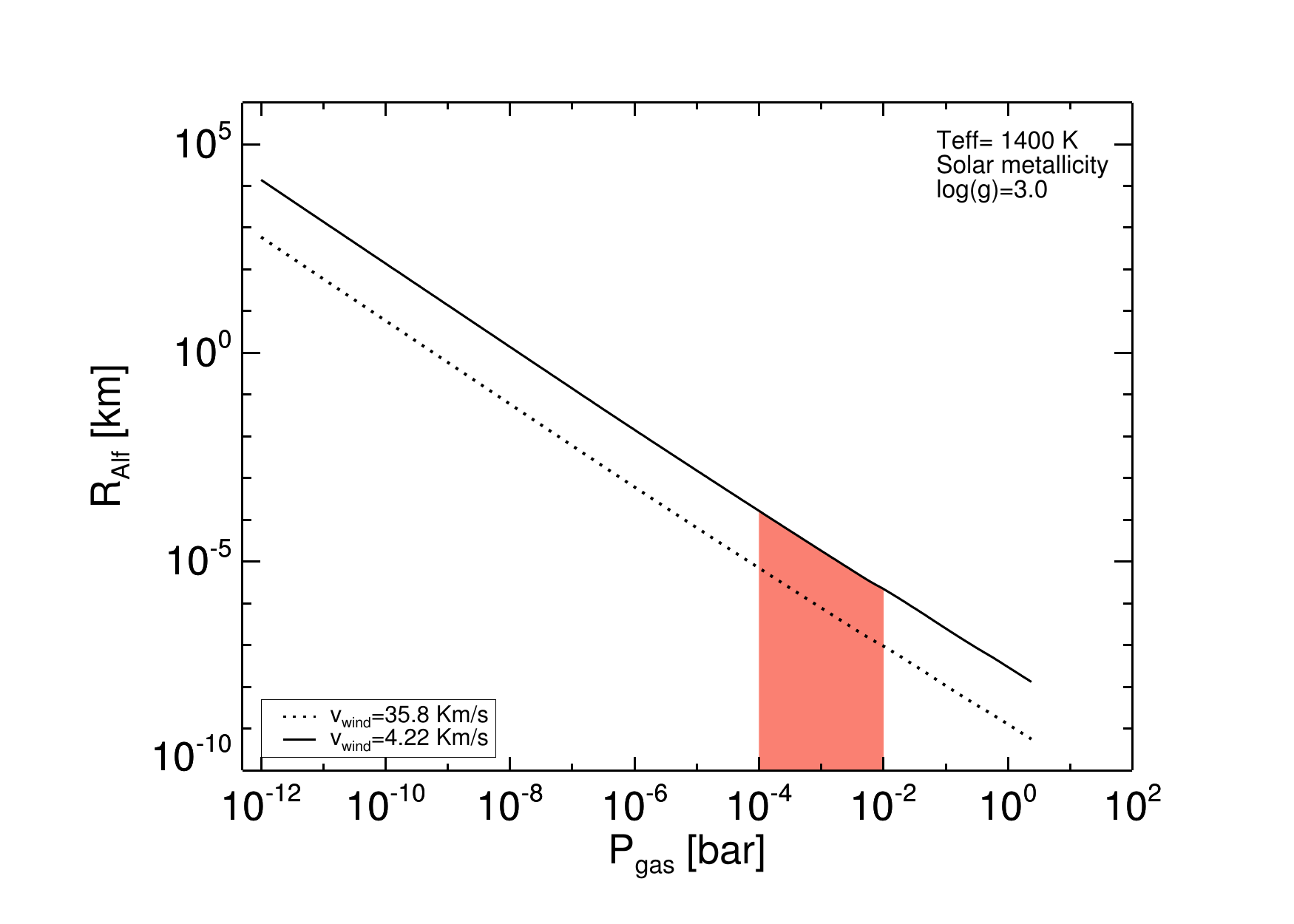}
\caption{Size of cloud holes, $R_{\rm Alf}(z)=v_{\rm wind}/\nu_{\rm
    ni}(z)$, that would be  caused by Alfv{\'e}n ionisation induced electrostatic
  cloud particle destruction if the respective hydrodynamic wind speeds can be reached. $R_{\rm Alf}$ is calculated for two
  cases: {\bf solid line:} wind speed reached values to ionise
  hydrogen ($v_{\rm wind}=38.5$ km\,s$^{-1}$, $n_{\rm gas}=n_{<H>}$, $v_{\rm th, i}=v_{\rm th, H^-}$),
  {\bf dotted line:} wind speed reached values to ionise K ($v_{\rm
    wind}=4.22$ km\,s$^{-1}$, $n_{\rm gas}=n_{<H>}$, $v_{\rm th, i}=v_{\rm th, K^+}$). The red-shaded
  area is the pressure range where dynamic atmosphere models reach
  maximum values of 5$\,\ldots\,$6 km\,s$^{-1}$. }
 \label{fig:Rralf}
\end{figure}

The timescale for Coulomb explosion to occur can be derived from
Eq. (\ref{eq:surface-solution}) using setting $N=N_{\rm ce}$ for a
given grain size (Eq.~\ref{coulomb}). This timescale is plotted as a
function of atmospheric pressure $p_{\rm gas}$ [dyn/cm$^{2}$] in
Figure \ref{fig:time-scale}, and the ionisation rate parameter is
explored.
%\sed{We predict that, if the gas-phase electrons are thermalized with the neutral gas-phase before making contact with the dust 
%grain, then Coulomb explosions will not occur anywhere in our model atmosphere}. As we stated above, it is unlikely that 
%electrons produced by an energetic process will have thermalized before colliding with a grain, so we examine the atmospheric region 
%in which coulomb explosions will take place for different electron temperatures, $T_e$ [eV]. We find that grains begin to be 
%electrostatically disrupted when $T_e \gtrsim 10$ eV \sed{($\sim 10^5$ K)}, but only for the upper atmosphere, 
%$p_{\rm gas} \lesssim 10^{-2}$ dyn/cm$^2$. For $T_e = 100$ eV \sed{($\sim 10^6$ K)}, Coulomb explosions will have a significant 
%effect on the size distribution of the dust for $p_{\rm gas} \lesssim 10^2$ dyn/cm$^2$.
The timescale for Coulomb explosion depends critically upon the
ionization rate. For fast electrons, when $T_e > 10$ eV ($\approx
10^5$K) , the timescale ranges from about 100 days for a weak source
of ionization, $\zeta = 10^{-17}$ s$^{-1}$. For cosmic rays as an
example of a weak ionisation source, the timescale for Coulomb
explosion drops to on the order of one day, and for strong ionizing
process, such as Alfv\'{e}n ionization, the timescale drops
significantly to between 1 $\mu$s to 1 ms. This broad range of
timescales means that the ionizing source has a significant effect on
the size distribution of grains in the upper atmosphere. This leads in
particular to the conclusion that the occurrence of highly efficient
local ionisation processes like Alfv\'{e}n ionization lead to a local
destruction of cloud particles through electrostatic effects.  
%We note, however, that Eq.~\ref{eq:tcrit-approx} was derived under the
%assumption of a weak ionisation source (small $\zeta$) and/or a very
%inefficient recombination (small $\alpha$).
 The resulting sputtering
products increase the number of small grains which will lead to a
local increase of cloud opacity. Each of these sputtering products may
carry different charges depending on their individual size. There is
no reason to believe that sputtering products are of the same size
similar to mechanical destruction processes (e.g. \citealt{guet2010}).
Cloudy atmosphere of brown dwarfs with strong winds that result in
Alfv\'{e}n ionization could therefore mimic the presence of star spots
or similar variability pattern if the affected area is large
enough. The 'dark spot' would only be present until the cloud
particles have grown to larger sizes so that no opacity contrast to
its surrounding will be detectable any more. The size of such spots
would be $R_{\rm Alf} \approx v_{\rm wind}/\nu_{\rm ni}$ assuming the
ion can follow the hydrodynamic wind. Figure~\ref{fig:Rralf}
demonstrate the potential sizes of such spots for the case of a fully
ionised molecular hydrogen gas (solid line) and the case of that only
potassium (K, dotted line) could be ionised. Simulations by
\cite{dobbs2013} show that a maximum wind speed of 5$\,\ldots\,$6
km\,s$^{-1}$ is reached in the pressure interval p$_{\rm
  gas}=10^{-4}\ldots\,10^{-2}$ bar (red shaded area) in an irradiated
giant gas planet atmosphere. Similar models for irradiated brown
dwarfs are not available yet. Brown dwarf circulation models for
non-irradiated but fast rotating objects achieve hydrodynamic
velocities in the atmosphere below the necessary Alfv\'{e}n ionization
threshold at present (\citealt{zhang2014}). In the case of a giant gas
planets were the local wind speed is sufficient inside the atmosphere
where p$_{\rm gas}=10^{-4}\ldots\,10^{-2}$ bar, the diameter of an
Alfv\'{e}n ionization induced cloud hole would be 0.1m at the lowest
pressure end of the interval, too small to make any observational
effect. If appropriate velocities occur at lower pressures, cloud-free
spots could be just 10m of size, still far below an observational
effect.

%For the lowest example of  $\zeta$, a
%$\Sigma_s = 10^7$ Pa, $T_{\rm e} = 500$ K, $n_{\rm d} = 10^{-10} n_{\rm gas}$, $n_{\rm gas} = 10^{10}$ cm$^{-3}$, $a = 0.01$ $\mu$m. 
%For these conditions, $\zeta < 4 \times 10^{-14}$ s$^{-1}$, which suggests that Eq. (\ref{eq:tcrit-approx}) is valid for all values 
%of $\zeta$ for cosmic rays, but probably would not be valid for Alfv\'{e}n ionization.

\section{Summary\label{summary}}

Thermal ionisation in brown dwarf and giant gas planet atmospheres with local temperatures between 100K$\,\ldots\,4000$K and high local gas pressures between $10^{-7}$bar$\,\ldots\, 5$bar allows for electrostatic interactions and magnetic coupling of such ultra-cool and chemically rich gases. Non-equilibrium processes like cosmic ray ionisation and discharge processes in clouds will increase the local pool of free electrons for a certain time which will, however, not increase the magnetic field coupling which is mainly determined by a rather strong magnetic field density in particular in brown dwarfs. Cosmic rays and lighting discharges do have a distinct effect on the composition of the local atmospheric gas. Cosmic rays affect the atmosphere through air showers which were modelled with a 3D Monte Carlo radiative transfer code to be able to visualise their spacial extent. Only the upper-most atmospheric layers  of extrasolar planets and brown dwarfs are effected by cosmic ray triggered air showers similar to solar system planets. 

Atmospheres of giant gas planets and brown dwarfs are so cold that
clouds made of mixed-material mineral particles that condense directly
from the gas phase. Gravitational settling (rain out) determines the
vertical extension and causes a large-scale charge separation of
charged cloud particles of different sizes. Cloud particles are
charged due to triboelectric processes in such highly dynamic
atmospheres. This dynamics is driven by irradiation or rotation which
links the planet and brown dwarf to its stellar and galactic
environment. If hydrodynamic winds achieve high enough local speeds,
Alv{\'e}n ionisation creates pockets of a high degree of
ionisation. The free electrons will interact with the cloud particles
and electrons will contribute to the charge accumulation on the grain
surface. We estimate that this process is too inefficient to cause an
electrostatic disruption of the cloud particles, hence, the cloud
particles will not be destroyed by Coulomb explosion for the local
temperature in the collisional dominated brown dwarf and giant gas
planet atmospheres.  If, however, the ionisation rate of the
atmosphere increases, Coulomb explosion may cause the emergence of
cloud holes. The potential size of such holes in an extended cloud
deck is too small and would occur to far inside the cloud to mimic the
effect of magnetic field induced star spots.

\section*{Acknowledgments}

%\bigskip
We highlight financial support of the European Community under the FP7
by the ERC starting grant 257431.

\footnotesize{
\bibliographystyle{mn2e}
\bibliography{bib}{}}

\end{document}